\documentclass[12pt]{article}

\usepackage{graphicx}
\usepackage{amssymb}
\usepackage{amsmath}

\numberwithin{equation}{section}

\begin{document}
\baselineskip=16pt
\begin{titlepage}
\begin{flushright}
{\small KYUSHU-HET-107}\\[-1mm]{\small KUNS-2074}%
\end{flushright}
\begin{center}
\vspace*{12mm}

{\Large\bf%
Holographic Construction of Technicolor Theory%
}\vspace*{10mm}

Takayuki Hirayama$^1$\footnote{e-mail: 
{\tt hirayama@th.physik.uni-bonn.de}} and 
Koichi Yoshioka$^2$\footnote{e-mail: 
{\tt yoshioka@gauge.scphys.kyoto-u.ac.jp}}
\vspace*{4mm}

${}^1$
{\it Physikalisches Institut der Universitaet Bonn,
Nussallee 12, 53115 Bonn, Germany}\\[1mm]
${}^2$
{\it Department of Physics, Kyushu University, Fukuoka 
812-8581, Japan}\\[1mm]
${}^2$
{\it Department of Physics, Kyoto University, Kyoto 606-8502, Japan}
\end{center}
\vspace*{10mm}

\begin{abstract}\noindent%
We construct a dual description of technicolor theory based on the
D4/D8 brane configuration. A strongly-coupled technicolor theory is
identified as the effective theory on D-branes, and from the
gauge/gravity correspondence, we explore the weakly-coupled
holographic description of dynamical electroweak symmetry
breaking. It is found from the D-brane probe action that the masses of
$W$ and $Z$ bosons are given by the decay constant of technipion, and
the technimesons become hierarchically heavy. Moreover, the couplings
of heavier modes to standard model fermions are rather suppressed. The
oblique correction parameters are also evaluated and found to be small
except for the $S$ parameter, which can be reduced by modifying the
model. The fermion fields are introduced at the intersections of
D-branes and their masses are generated via massive gauge bosons from
open strings stretching between D-branes.
\end{abstract}

\end{titlepage}

\newpage

\section{Introduction}

Future particle experiments such as the Large Hadron Collider (LHC)
will provide important data on the sector of electroweak gauge
symmetry breaking. In the Standard Model (SM), the elementary scalar
fields, the Higgs bosons, are responsible for the symmetry breaking,
though there is a well-known problem of gauge hierarchy between the
Planck and electroweak scales. One of the alternatives to elementary
Higgs is the dynamical electroweak symmetry breaking induced by a
strongly interacting gauge theory, the technicolor
scenario~\cite{TC}. It has been known, however, that technicolor
models often suffer from the difficulty of passing the electroweak
precision tests through the oblique corrections~\cite{oblique}. Since
a strongly interacting dynamics is involved in the analysis, it is
still an important issue whether a realistic technicolor model can be
constructed.

The recent development of brane physics in string theory provides us
an alternative way to analyze strong coupling region of gauge theory
via weakly-coupled gravitational description. The original proposal of
the gauge/gravity correspondence claims that the supergravity on
AdS$_5\times S^5$ is dual to the four-dimensional ${\cal N}=4$ $SU(N)$
super Yang-Mills theory with large $N$ and large 't~Hooft
coupling~\cite{AdSCFT,AdSCFT2}, and various derivatives have been
discussed in the literature. In particular, there have been attempts
to construct a holographic description of Quantum Chromo Dynamics
(QCD), which is a strongly interacting gauge theory at low energy
regime. Among them the model of Ref.~\cite{SS} realizes the
non-abelian chiral symmetry breaking from the D-brane geometry and
predicts the vector meson mass spectrum and interactions which are
comparable with the experimental data.

The dynamical electroweak symmetry breaking in the technicolor
scenario is regarded as a scale-up version of chiral symmetry breaking
in QCD\@. In this paper, we construct a dual description of
technicolor by applying the holographic gauge/gravity
correspondence. Since the holographic description is in the weakly
coupling regime, it enables us to treat the
non-perturbative dynamics of technicolor theory in a perturbative
way. Furthermore, a dual technicolor theory is constructed from the
D-brane configuration and the gauge/gravity correspondence makes it
possible to analyze the technicolor dynamics in quantitative
treatment. In order to gauge the flavor chiral symmetry in QCD, it is
assumed that the six-dimensional extra space in string theory are
compactified. In the original gauge/gravity correspondence, this
procedure is expected to introducing a cutoff near the AdS boundary
and giving appropriate boundary conditions at the cutoff. From the
holographic description, we can calculate the strength of gauge
couplings and the mass spectra of SM gauge bosons and composites
fields which are analogous to QCD-like mesons in the technicolor
theory. The gauge bosons other than the SM ones are shown to become
hierarchically heavy. We also discuss how to introduce SM quarks and
leptons into our scheme and compute their minimal couplings to the SM
gauge bosons and heavier modes. The fermion masses are 
induced by a similar mechanism to the extended technicolor
theory~\cite{ETC}. The oblique correction parameters are explicitly
calculated and are found to be small except for the $S$ parameter. We
comment on possibilities to suppress the $S$ parameter in our model.

This paper is organized as follows. In the next section we describe
the D-brane configuration to define our technicolor theory as the
effective theory on the D-branes. In Section~\ref{sec:DSB}, its
holographic dual description is explored where the probe branes
describe the action below the scale of techniquark
condensation. Solving the equations of motion both approximately and
numerically, we show how the SM gauge bosons and composite fields are
described and evaluate their masses and
interactions. Section~\ref{sec:fl} discusses an idea of introducing SM
matter fields by utilizing additional D-branes. We derive the
Lagrangian for SM matter fields from the holographic description and
estimate their masses and gauge interaction strength. In
Section~\ref{sec:STU} we examine whether the model passes the
electroweak precision tests by evaluating oblique corrections to the
electroweak observables. Some comparison with the so-called higgsless
models~\cite{higgsless} is mentioned in
Section~\ref{sec:comp}. Finally we conclude and discuss open issues in
the last section.

\bigskip
\section{D-brane Configuration: The Gauge Sector}

\begin{figure}[t]
\begin{center}
\includegraphics[width=8.2cm]{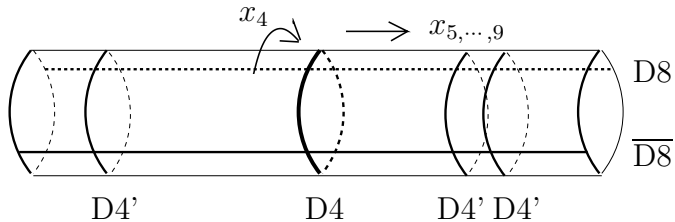}
\put(-146,75){$x_4$}
\put(-74,70){$x_{5,\cdots,9}$}
\put(-121,0){D4}
\put(-202,0){D4'}
\put(-71,0){D4'}
\put(-50,0){D4'}
\put(3,52){D8}
\put(3,20){$\overline{\rm D8}$}\smallskip
\caption{The D-brane configuration near the technicolor D4-branes in
the flat space. The extra dimensions transverse to D4-branes
($x_{5,\cdots,9}$) are assumed to be compactified. (D4' denotes
possible locations of flavor D4-branes for SM matter fields, which
will be explained later.)\bigskip}
\label{fig:brane}
\end{center}
\end{figure}

In this section we describe the D-brane configuration in the flat
space background of type IIA string theory which realizes a
technicolor scenario as the effective theory on the D-branes. The
configuration consists of D4, D8 and anti-D8 ($\overline{\rm D8}$)
branes. The coincident $N_{TC}$ D4-branes realizes pure $SU(N_{TC})$
Yang-Mills theory\footnote{The overall $U(1)$ factor decouples.} in
compactifying one spacial direction on a circle $S^1$ with the
anti-periodic boundary condition of the fermionic variable on the
D4-branes. The boundary condition leads to the fermion zero mode being
projected out, and the scalar modes become massive due to
supersymmetry-breaking quantum effects. We thus identify this
$SU(N_{TC})$ as the technicolor gauge symmetry. In this work we refer
to $N_{TC}$ D4-branes as the technicolor branes. The techniquarks are
provided by introducing $N_f$ sets of D8 and 
$\overline{\rm D8}$-branes. They are localized at different (possibly
opposite) points in the $S^1$ direction and the open string stretching
between the technicolor D4 and D8 ($\overline{\rm D8}$) branes
provides a four-dimensional massless chiral (anti-chiral) fermion,
i.e.\ a pair of techniquarks, as the lowest massless mode. The cartoon
of D-brane configuration is shown in Fig.~\ref{fig:brane}.
The five-dimensional transversal directions to the technicolor
D4-branes are assumed to be compactified in order to have a finite
Newton constant in four dimensions. Then the gauge fields on the
D8-branes are dynamical in four dimensions and induce $U(N_f)_L$ gauge
symmetry. Similarly we have $U(N_f)_R$ gauge symmetry on the
$\overline{\rm D8}$-branes. The left (right) handed techniquark from
D4-D8 (D4-$\overline{\rm D8}$) branes is bi-fundamentally charged
under $SU(N_{TC})\times U(N_f)_L\,$ [$SU(N_{TC})\times U(N_f)_R$].
\begin{align*}
\begin{array}{c|ccc}
  & SU(N_{TC}) & U(N_f)_L & U(N_f)_R \\ \hline
  Q_L & \Box & \Box & \\
  Q_R & \Box & & \Box
\end{array}
\end{align*}
where the blanks denote singlet representations hereafter.
The $U(N_f)_L\times U(N_f)_R$ symmetry is expected to be dynamically
broken by the condensation of 
techniquarks $\langle\overline Q_RQ_L\rangle$ and thus the electroweak
gauge symmetry is embed in $U(N_f)_L\times U(N_f)_R$. In addition to
the gauge fields, there are scalar and spinor modes on the D8 
and $\overline{\rm D8}$-branes. These fields are assumed to receive
loop-induced masses since supersymmetry is broken and their mass terms
are not prohibited by any symmetry. In the $\alpha'\to0$ limit, there
are no tachyonic states, but for a finite $\alpha'$ there is an
instability caused by the closed-string exchange between D8 
and $\overline{\rm D8}$-branes. The mode associated with this
instability could are stabilized by some mechanism in string theory
such as fluxes, Casimir effects or non-perturbative effects. Later,
additional D4-branes which provide quarks and leptons are introduced
(D4' in Fig.~\ref{fig:brane}) and the positions of D4'-branes are the
moduli which should also be stabilized. It is assumed that
Ramond-Ramond charges and the cosmological constant are both canceled
by properly introducing D-branes, anti D-branes or orientifolds away
from the technicolor branes.

Depending on how to embed the electroweak gauge symmetry 
in $U(N_f)_L\times U(N_f)_R$, different types of phenomenological
models can be constructed. In this paper we investigate the simplest
choice where the number of D8-branes is minimal, i.e.\ $N_f=2$. In
this case there are two ways to realize the electroweak symmetry. The
first choice is identifying $SU(2)_L\subset U(2)_L$ as the electroweak
$SU(2)$ symmetry in the SM and $U(1)\subset U(2)_R$ as the hypercharge
$U(1)_Y$. The other is the identification 
that $SU(2)\times U(1)\subset U(2)_L$ is the electroweak symmetry and
$U(1)\subset U(2)_R$ is an extra $U(1)$ which is used to realize the
desired symmetry breaking pattern. While both choices of embedding are
worth consideration, in this paper we investigate the first pattern of
electroweak symmetry breaking.

The overall $U(1)$'s in $U(2)_L\times U(2)_R$ are related to the
positions of D8 and $\overline{\rm D8}$-branes and supposed to be
broken. To have the electroweak symmetry, $SU(2)_R\subset U(2)_R$ is
broken down to $U(1)_Y$ with an adjoint Higgs field at a high
scale. Notice that there is an adjoint scalar field on 
the $\overline{\rm D8}$-branes which can play as the Higgs field
inducing such breaking. Thus the viable gauge symmetry becomes 
$SU(2)_L\times U(1)_Y$ under which the techniquarks have the following
quantum charges:
\begin{center}
$\begin{array}{c|ccc}
  & SU(N_{TC}) & SU(2)_L & U(1)_Y \\ \hline
  Q_L & \Box & \Box & \\
  Q_R & \Box & & (\frac{1}{2},\frac{-1}{2})
\end{array}$\medskip
\end{center}
Below the technicolor scale $\Lambda_{TC}$, at which the gauge
coupling of technicolor gauge theory becomes strong, the techniquarks
are expected to be condensed, 
i.e.\ $\langle\bar{Q}_L^\alpha Q_{R\beta}^{}\rangle\sim 
N_{TC}\Lambda_{TC}^3\delta^\alpha_\beta\,$ ($\alpha,\beta =1,2$), and
the dynamical electroweak symmetry breaking is realized
\begin{align}
  SU(2)_L\times U(1)_Y \;\rightarrow\; U(1)_{EM}.
\end{align}

\bigskip
\section{Holographic Dual Description of Technicolor}
\label{sec:DSB}

The holographic dual description of the technicolor theory given above
is obtained from the gauge/gravity correspondence, that is, by
replacing the technicolor D4-branes with their near horizon
geometry. The near horizon geometry of D4-branes compactified on $S^1$
with supersymmetry-breaking boundary condition~\cite{D4metric} is
\begin{gather}
  ds^2 \;=\; \left(\frac{u}{R}\right)^\frac{3}{2}
  \big(dx_\mu^2+f(u)dx_4^2\big)
  +\left(\frac{R}{u}\right)^\frac{3}{2}
  \left(\frac{du^2}{f(u)}+u^2d\Omega_4^2\right),
  \\[2mm]
  f(u) \,=\, 1-\frac{u_K^3}{u^3}, \hspace{5ex}
  R^3 \,=\, \pi g_s N_{TC} l_s^3, \nonumber
\end{gather}
and the dilaton $\phi$ and the Ramond-Ramond four-form field strength
$F_4$ are given by
\begin{align}
  e^\phi \,=\, g_s\left(\frac{u}{R}\right)^\frac{3}{4}, \hspace{5ex}
  F_4 \,=\, \frac{2\pi N_{TC}}{V_4}\epsilon_4.
\end{align}
The D4-branes extend to the four-dimensional spacetime 
$x_\mu$ ($\mu=0,1,2,3$) and the $x_4$ direction which is compactified
on a circle $S^1$ with the radius $(M_K)^{-1}$:
\begin{align}
  x_4 \,\sim\, x_4 +\frac{2\pi}{M_K}, \hspace{5ex}
  M_K \,=\, \frac{3u_K^{\frac{1}{2}}}{2R^{\frac{3}{2}}} .
\end{align}
The coordinate $u$ is a radial direction transversal to the D4-branes,
and $d\Omega_4^2$, $V_4$ and $\epsilon_4$ are the metric, volume and
line element of the unit four-dimensional sphere. The constant
parameter $R$ is proportional to the number of D4-branes $N_{TC}$. The
technicolor gauge coupling $g_{TC}^{}$ at the compactification scale
$M_K$ is determined by the string coupling $g_s$ and the string length
$l_s=\alpha'^{1/2}$ and is given by $g_{TC}^2=2\pi g_sl_sM_K$. The
holographic dual description is valid in the region
$1\ll N_{TC}g_{TC}^2\ll g_{TC}^{-4}$~\cite{stringQCD}.

The existence of technicolor D4-branes modifies the geometry near
themselves, that is, a throat is developed. Therefore we are looking
at the geometry of throat. This geometry is trustful in the region
$R\gg u\,$ ($\geq u_K$) and the contribution to the four-dimensional
Planck constant is negligible compared with the bulk contribution as
long as the throat volume is much smaller than the bulk one without
the throat. If one supposes that the size of compactification is the
same order of the string length $l_s$ except for the $x_4$ direction,
the volume of six-dimensional extra space is $2\pi l_s^5/M_K$. Then as
long as the maximal value of $u$ is smaller 
than $R$ ($u_{\rm max}\ll R N_{TC}^{-5/6}$), the throat has a
suppressed size, as desired.

\begin{figure}[t]
\begin{center}
\includegraphics[width=9cm]{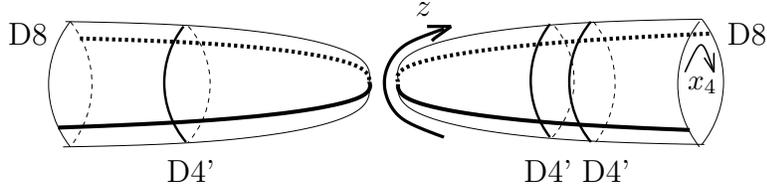}
\put(-211,-13){D4'}
\put(-54,-13){D4'}
\put(-76,-13){D4'}
\put(1,39){D8}
\put(-271,39){D8}
\put(-117,50){$z$}
\put(-14,22){\small $x_4$}\medskip
\caption{The D8-brane configuration in the near horizon geometry. A
pair of original D8 and $\overline{\rm D8}$-branes are smoothly
combined into a single D8-brane.\bigskip}
\label{fig:D4bg}
\end{center}
\end{figure}

Since a large value of $N_{TC}\gg N_f$ ($=2$) is taken for the
validity of holographic description, the D8 and $\overline{\rm D8}$
branes are treated as probes in the D4 geometry. In the flat space the
D8-branes are localized at a constant $x_4$ and still reside in the
same point on the curved geometry since the metric coefficients do not
explicitly depend on $x_4$. Then the coefficient $f(u)$ of $dx_4^2$
goes to zero at $u=u_K$, and the D8-branes are smoothly connected with
the $\overline{\rm D8}$-branes at this point and make up smooth
D8-branes, see Fig.~\ref{fig:D4bg}.
In the holographic dual description of technicolor theory with a large
't~Hooft coupling, the connection of D8 and $\overline{\rm D8}$-branes
is interpreted as the dynamical breaking of $U(N_f)_L\times U(N_f)_R$
symmetry to the diagonal one $U(N_f)$. This is because only the
simultaneous rotation of D8 and $\overline{\rm D8}$-branes remains
intact. In Ref.~\cite{SS}, the smoothly connected D8-brane describes
the chiral symmetry breaking in QCD and provides the meson spectrum
and interactions. In the present model ($N_f=2$), the D8-branes action
describes the dynamical electroweak symmetry breaking in technicolor
theory, and provides the SM gauge bosons (the photon, $W$ and $Z$
bosons) as well as technimesons below the scale of the techniquark
condensation.

The probe D8-brane action is given by the Dirac-Born-Infeld action in
the curved geometry. We focus on the gauge sector while scalars and
spinors on the branes may become massive due to high-scale
supersymmetry breaking. The relevant action up to the quadratic level
is obtained from the Yang-Mills approximation of the Dirac-Born-Infeld
action
\begin{align}
  S \;=\; -\tau(2\pi\alpha')^2\int d^9 x\; e^{-\phi}\sqrt{-g}\,
  \mbox{Tr} g^{ac}g^{bd}F_{ab}F_{cd} ,
\end{align}
where $\tau=(2\pi)^{-8}l_s^{-9}$ is the tension of 
D8-brane, $g_{ab}$ ($a,b=0,\cdots,8$) is the induced metric and
$F_{ab}$ is the field strength of $U(2)$ gauge fields on the probe
D8-branes. It is noted that there exists a single $U(2)$ gauge theory
on the connected D8-branes. The D8-branes are localized at $x_4=0$ and
the induced metric is given by
\begin{gather}
  \quad ds^2 \;=\; 
  \left(\frac{u_K}{R}\right)^\frac{3}{2} K(z)^\frac{1}{2}dx_\mu^2
  +\left(\frac{R}{u_K}\right)^\frac{3}{2} K(z)^\frac{-1}{2}
  \left(\frac{4}{9} u_K^2 K(z)^\frac{-1}{3} dz^2
    +u_K^2 K(z)^\frac{2}{3} d\Omega_4^2\right),
  \\[3mm]
  u^3 \,\equiv\, u_K^3 K(z), \hspace{7ex}
  K(z) \,\equiv\, 1+z^2,
\end{gather}
where we have defined a new dimensionless coordinate $z$ which goes
along the D8-brane ($-\infty<z<\infty$)\footnote{%
$\big(u(-z),x_4\big)=\big(u(z),x_4+\pi/M_K\big)$ and 
$|x_4|\leq\pi/M_K$ in the new coordinates system.}.
One may understand that the D8 ($\overline{\rm D8}$) branes are
described in the $z>0$ ($z<0$) region and they are smoothly connected
with each other at $z=0$. The four-dimensional gauge action below the
compactification scale is obtained by integrating over extra five
dimensions (the $z$ and $S^4$ directions). The metric has the $SO(5)$
invariance of $S^4$ and Kaluza-Klein modes in the compactification are
parameterized by angular momenta along $S^4$. The nonzero momentum
modes are heavy $\gtrsim M_K$ and in the following discussion only the
zero momentum modes may be relevant. We therefore focus on
$SO(5)$-invariant modes and evaluate the $S^4$ integration to obtain
the five-dimensional effective action
\begin{gather}
  S \;=\; -\int d^4x \int_{-z_R^{}}^{z_L^{}}\!dz\; \mbox{Tr}\left[
    \frac{1}{4}K(z)^\frac{-1}{3} F_{\mu\nu}^2
    +\frac{M_K^2}{2}K(z)F_{\mu z}^2 \right],\;\;
  \\[2mm]
  F_{ab} \,=\, \partial_aA_b-\partial_bA_a-ig_5[A_a,A_b],
  \hspace{5ex}
  g_5^{-2} =\, \frac{2}{3}k^2R^\frac{9}{2}u_K^\frac{1}{2}
  \tau V_4g_s^{-1}(2\pi\alpha')^2,
  \label{g5}
\end{gather}
where the Lorentz indices $\mu$, $\nu$ are contracted by the
four-dimensional Minkowski metric hereafter. Here we have dropped
gauge fields along the $S^4$ directions which are expected to obtain
masses from supersymmetry breaking and compactification. The gauge
boson $A_\mu$ has been rescaled such that the coefficient of kinetic
term is properly $1/4$ except for the $K(z)$ factor. As mentioned in
the introduction, the boundaries of extra dimension have been
introduced at $z_L^{}\;(>0)$ and $-z_R^{}\;(<0)$, and the bulk
geometry outside the throat is integrated out. The parameters $z_L^{}$
and $z_R^{}$ reflect the volume of D8 and $\overline{\rm D8}$-branes
in the extra dimension, and the four-dimensional gauge couplings of
$U(N_f)_{L,R}$ are inversely proportional to $z_{L,R}^{}$, as seen
below. We have also introduced a parameter $k$ in the definition of
gauge coupling $g_5$ which represents how the D8-branes extend in the
bulk.

In order to have the four-dimensional effective action integrating
over the $z$ direction, one needs to specify the boundary conditions
of gauge fields at $z=z_L^{}$ and $z=-z_R^{}$. Since the $z>0$ ($z<0$)
region is understood as the D8 ($\overline{\rm D8}$) branes on which
the $SU(2)_L$ ($U(1)_Y$) gauge symmetry is realized, it is found that
the following conditions are appropriate for the present situation
($+$/$-$ denotes the Neumann/Dirichlet boundary condition):
\begin{alignat}{2}
  A_\mu^{1,2} (z_L^{}) \;:&\;\; +\,& \hspace{12ex} 
  A_\mu^{1,2} (-z_R^{}) \;:&\;\; -\,
  \nonumber \\
  A^3_\mu(z_L^{}) \;:&\;\; +\, & \hspace{12ex}
  A^3_\mu(-z_R^{}) \;:&\;\; +\,
  \\
  A^4_\mu (z_L^{}) \;:&\;\; -\, & \hspace{12ex}
  A^4_\mu (-z_R^{}) \;:&\;\; -\, \nonumber
\end{alignat}
where $A^{1,2,3}_\mu$ and $A^4_\mu$ are the $SU(2)$ and $U(1)$ gauge
fields, respectively. We take the $A_z=0$ gauge hereafter. In this
gauge, the scalar zero modes in $A_\mu$ are taken into account. While
$A^4_\mu$ has such a scalar mode which is interpreted as the
Nambu-Goldstone boson associated with the global axial $U(1)_A$
symmetry, there is nonzero mixed gauge anomaly of 
$SU(N_{TC})^2\times U(1)_A$ and the scalar zero mode becomes massive
due to the Green-Schwarz mechanism. A Dirichlet boundary condition
may be interpreted as taking a scalar expectation value which causes
symmetry breaking infinite.

Expanding the gauge fields with orthonormal wavefunctions as
$A_\mu(x,z)=\sum_n A^{(n)}_\mu(x) \psi_n(z)$, we obtain the equations
of motion for the eigenmodes from the above action
\begin{align}
  \partial_z^2\psi_n(z) \;=\; \frac{-2z}{1+z^2}\partial_z\psi_n(z)
  -\lambda_n K(z)^\frac{-4}{3}\psi_n(z) ,
  \label{eom}
\end{align}
with the normalization condition
\begin{align}
  \int_{-z_R^{}}^{z_L^{}}dz\, K(z)^\frac{-1}{3} \psi_n(z)^2 \,=\, 1.
\end{align}
The mass of the eigenmode $A^{(n)}_\mu$ is given by 
$m_n^2=\lambda_nM_K^2$. The zero-mode wavefunctions are easily found
and are proportional to
\begin{align}
  \psi_{0L}(z) \,=\, \frac{1}{2}+\frac{\arctan(z)}{\pi}, \hspace{7ex}
  \psi_{0R}(z) \,=\, \frac{1}{2}-\frac{\arctan(z)}{\pi}.
\end{align}
The existence of two massless modes inherits the fact that there are
originally two gauge sectors $U(2)_{L,R}$. The wavefunctions
$\psi_{0L}(z)$ and $\psi_{0R}(z)$ are localized in the positive and
negative $z$ region respectively and then correspond to 
the wavefunctions of $U(2)_L$
and $U(2)_R$ gauge fields in the technicolor side. Furthermore,
$\psi_{0L}$ ($\psi_{0R}$) becomes normalizable as long as $z_L^{}$
($z_R^{}$) is finite. This is consistent with the facts that $z_L^{}$
($z_R^{}$) reflects the volume of D8 ($\overline{\rm D8}$) branes and
the gauge fields on the D8 ($\overline{\rm D8}$) branes become
dynamical in four dimensions if the volume of D8 ($\overline{\rm D8}$)
is finite along the extra dimensions.

Solving the equations of motion \eqref{eom}, we find that the
Kaluza-Klein decompositions of gauge fields take the following forms:
\begin{align}
  A_\mu^{\alpha}(x,z) \,=&\; W_\mu^{\alpha}(x)\psi_W(z) 
  +\sum_{n=2} X_\mu^{\alpha(n)}(x)\psi^\alpha_n(z),
  \hspace{5ex}  (\alpha=1,2)
  \\
  A_\mu^3(x,z) \,=&\; Q_\mu(x)\psi_Q +Z_\mu(x) \psi_Z(z) 
  +\sum_{n=2} X_\mu^{3(n)}(x)\psi^3_n(z),
  \\
  A^4_\mu(x,z) \,=&\; \sum_{n=1} X_\mu^{4(n)}(x)\psi^4_n(z).
\end{align}
For the boundaries far away from the origin ($z_{L,R}^{}\gg1$), the
lower mode wavefunctions are approximately given by 
\begin{align}
  \psi_Q \;\simeq&\; 
  \frac{1}{\sqrt{3(z_L^\frac{1}{3}+z_R^\frac{1}{3})}},
  \\
  \psi_Z(z) \,\simeq&\;
  \frac{1}{\sqrt{3(z_L^\frac{1}{3}+z_L^\frac{2}{3}z_R^\frac{-1}{3})}}
  \Big[\psi_{0L}(z) -z_L^\frac{1}{3}z_R^\frac{-1}{3}\psi_{0R}(z)\Big],
  \\[2mm]
  \psi_W(z) \,\simeq&\; 
  \frac{1}{\sqrt{3}}\,z_L^\frac{-1}{6}\psi_{0L}(z),
\end{align}
and the mass eigenvalues are
\begin{align}
  m_Q^2 \,=\, 0, \hspace{5ex}
  m_Z^2 \,\simeq\, \rho_Z^{}(z_L^\frac{-1}{3}+z_R^\frac{-1}{3}) M_K^2 ,
  \hspace{5ex}
  m_W^2 \,\simeq\, \rho_W^{} z_L^\frac{-1}{3}M_K^2,
  \label{mQZW}
\end{align}
where $\rho_Z^{}\simeq\rho_W^{}\simeq 0.11$, roughly independent of
$z_{L,R}^{}$. It is interesting to notice that the masses of other
Kaluza-Klein excited gauge bosons become $m_{X^{(n)}}^2\gtrsim M_K^2$
and hierarchically larger than the SM gauge boson 
eigenvalues \eqref{mQZW}. These formulas are found to well fit the
numerical results within a few percent errors in calculating the gauge
coupling constants in later sections.

\begin{figure}[t]
\begin{center}
\begin{minipage}{7.5cm}
\begin{center}
\includegraphics[width=6cm]{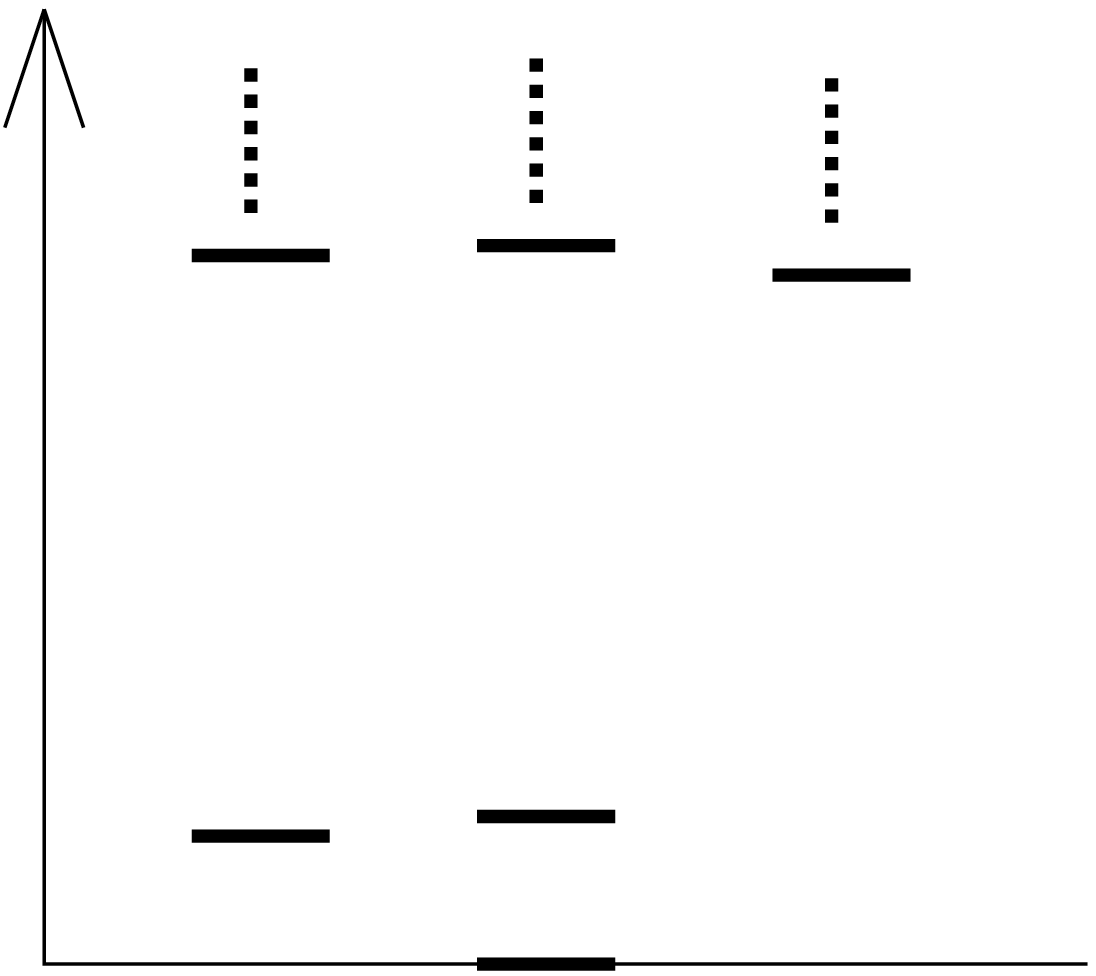}
\put(-138,-15){$A_\mu^{1,2}$}
\put(-92,-15){$A_\mu^3$}
\put(-46,-15){$A_\mu^4$}
\put(-170,158){$m$}
\put(-186,-0){$m_Q^{}$}
\put(-186,18){$m_W^{}$}
\put(-186,28){$m_Z^{}$}
\put(-186,109){$m_X^{}$}\medskip
\caption{A schematic picture of the Kaluza-Klein mass spectrum of
four-dimensional gauge bosons (technimesons).}
\label{fig:kkmass}
\end{center}
\end{minipage}
\hspace{4ex}
\begin{minipage}{7.65cm}
\begin{center}\hspace*{-4ex}
\includegraphics[width=8.6cm]{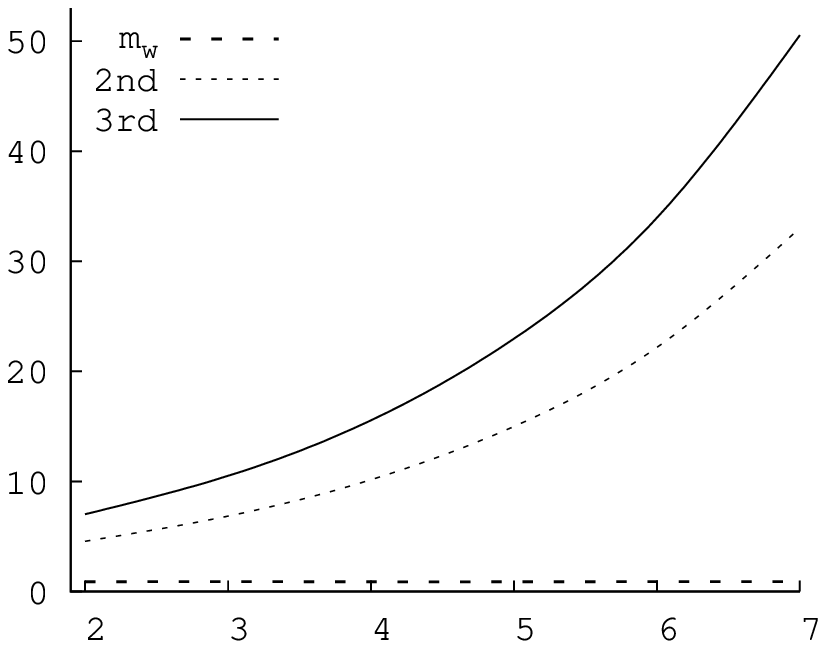}
\put(-240,166){{\small $m/m_Z^{}$}}
\put(-124,-6){\small $\log_{10}z_L^{}$}
\caption{The mass hierarchy among the weak bosons and 2nd and 3rd
Kaluza-Klein modes in $A^{1,2}_\mu$. The Weinberg angle is 
fixed.\bigskip} 
\label{fig:spectrum}
\end{center}
\end{minipage}
\end{center}
\end{figure}

We have found that there are four light gauge bosons $Q_\mu$, $Z_\mu$
and $W_\mu^{1,2}$ in addition to an infinite number of heavy
Kaluza-Klein modes $X_\mu^{\alpha(n)}$ ($\alpha=1,2,3,4$). For
a schematic pattern of mass spectrum, see Fig.~\ref{fig:kkmass}.
The wavefunction of the massless mode $Q_\mu$ is 
constant ($z$ independent), which is exactly given by 
$\psi_{0L}+\psi_{0R}$, and is understood as the unbroken $U(1)$ gauge
boson, the photon. With the formulas of wavefunctions, we can rewrite
the third-component gauge boson $A_\mu^3$ as
\begin{align}
  A_\mu^3(x,z) \;\simeq&\;\,
  \frac{z_L^\frac{-1}{6}}{\sqrt{3}}\Big[
  \big\{s_W^{}Q_\mu(x)+c_W^{}Z_\mu(x)\big\}\psi_{0L}(z)
  +\big\{c_W^{}Q_\mu(x)-s_W^{}Z_\mu(x)\big\}
  \frac{s_W^{}}{c_W^{}}\psi_{0R}(z)\Big]  \nonumber \\[2mm]
  & \hspace*{7ex} +\sum_{n=2} X_\mu^{3(n)}(x)\psi^3_n(z),
  \label{A3}
\end{align}
where we have introduced $\theta_W$
\begin{alignat}{2}
  \sin^2\theta_W \,\equiv\, s_W^2 \,=&\;
  \frac{z_L^\frac{1}{3}}{z_L^\frac{1}{3}+z_R^\frac{1}{3}},
  &\hspace{8ex}
  \cos^2\theta_W \,\equiv\, c_W^2 \,=&\; 
  \frac{z_R^\frac{1}{3}}{z_L^\frac{1}{3}+z_R^\frac{1}{3}}.
  \label{wa}
\end{alignat}
As explained before, the wavefunctions $\psi_{0L}$ and $\psi_{0R}$
respectively correspond to $U(2)_L$ and $U(2)_R$ in which the
electroweak gauge symmetry is contained as $SU(2)_L\subset U(2)_L$ and
$U(1)_Y\subset U(2)_R$. Therefore the expression \eqref{A3} indicates
that $Z_\mu(x)$ is interpreted as the $Z$ boson in the SM with the
identification that $\theta_W$ is the Weinberg angle. It is noted here
that the photon and the $Z$ boson are unified into a single gauge
field on the connected D8 branes in the present model. Interestingly
enough, the identification of the Weinberg angle \eqref{wa} is
consistent with the prediction of mass 
spectrum \eqref{mQZW}, i.e.\ the relation $m_W^2=m_Z^2\cos^2\theta_W$
is indeed satisfied. This fact confirms that $W_\mu^{1,2}(x)$
correspond to the $W$ bosons in the SM.

Substituting the Kaluza-Klein decomposition in the five-dimensional
action, we obtain the four-dimensional effective theory of the gauge
sector
\begin{align}
  S \,=& \int d^4x\,\bigg[
  -\frac{1}{4}(F_{\mu\nu}^Q)^2 -\frac{1}{4}(F_{\mu\nu}^Z)^2
  -\frac{1}{2} |F_{\mu\nu}^W|^2
  +\frac{1}{2}m_Z^2 Z_\mu^2 +m_W^2 |W_\mu|^2
  \nonumber \\[1mm]
  &\quad
  -i(eF_{\mu\nu}^Q+g_{WWZ}^{}c_W^{} F_{\mu\nu}^Z)W_\mu W_{\nu}^\dagger
  -\frac{i}{2}(eQ_\mu +g_{WWZ}^{}c_W^{} Z_\mu) 
  (W_\nu^\dagger\partial_\mu W_\mu +W_\nu\partial_\mu W_\mu^\dagger)
  \nonumber \\[2mm]
  &\quad
  +e^2{\cal O}_1(Q^2,W^2) +eg_{WWZ}^{}c_W^{}{\cal O}_2(Q,Z,W^2)
  +g_{WWZZ}^2c_W^2{\cal O}_3(Z^2,W^2) +g_{WWWW}^2{\cal O}_4(W^4)
  \nonumber \\[2mm]
  &\qquad
  -\sum_{a=\alpha,\,n}\Big\{
  \frac{1}{4}\big(F_{\mu\nu}^{X^a}\big)^2 
  +\frac{1}{2} m_{X^a}^2(X_\mu^a)^2 
  +(\mbox{interactions})\Big\}\,\bigg],
  \label{gaugeS}
\end{align}
where $F_{\mu\nu}^X=\partial_\mu X_\nu-\partial_\nu X_\mu$ ($X=Q,Z,W$)
and $W_\mu=(W_\mu^1-iW_\mu^2)/\sqrt{2}$. We have not written down the
four-point gauge interaction operators ${\cal O}_{1,2,3,4}$
explicitly. The strengths of self gauge interactions among the
electroweak gauge bosons are determined by the wavefunction profiles
\begin{align}
  e \,\equiv&\;\; g_5 \int dz\, K(z)^\frac{-1}{3} \psi_Q \psi_W(z)^2 
  \;=\, g_5 \psi_Q,
  \label{e-gauge} \\
  g_{WWZ}^{} \,\equiv&\;\; g_5c_W^{-1} \int dz\,
  K(z)^\frac{-1}{3} \psi_W(z)^2\psi_Z(z),
  \\
  g_{WWZZ}^2 \,\equiv&\;\; g_5^2c_W^{-2} \int dz\,
  K(z)^\frac{-1}{3} \psi_W(z)^2\psi_Z(z)^2,
  \\
  g_{WWWW}^2 \,\equiv&\;\; g_5^2\int dz\, K(z)^\frac{-1}{3}\psi_W(z)^4.
\end{align}
Since the wavefunctions are almost constant except for the small
$|z|$ region and $\psi_W(z)$ quickly goes to zero for negative $z$,
the following approximations hold: 
$e\simeq g_5c_W^{}z_R^{-1/6}/\sqrt{3}$ and
$g_{WWZ}^{}\simeq g_{WWZZ}^{}\simeq g_{WWWW}^{}\simeq
g_5z_L^{-1/6}/\sqrt{3}$. Thus the electroweak gauge couplings for
$SU(2)$ ($g$) and $U(1)_Y$ ($g'$) are found
\begin{alignat}{2}
  g \,\simeq&\; \frac{g_5}{\sqrt{3}z_L^\frac{1}{6}},
  &\hspace{10ex}
  g' \,\simeq&\; \frac{g_5}{\sqrt{3}z_R^\frac{1}{6}}.
 \label{u1yg}
\end{alignat}
Therefore it is again consistently understood that $z_L^{}$ and
$z_R^{}$ represent the volumes of D8 and $\overline{\rm D8}$ branes,
respectively. We will demonstrate in later section the numerical
results of mass spectrum and how $g_{WWZ}^{}$ etc.\ are close to the
$SU(2)$ weak gauge coupling which is determined from the fermion
vertices.

Let us turn to discuss the heavier gauge bosons. The lightest modes
among $X_\mu^{\alpha(n)}$ ($\alpha=1,2,3$) comes from the 2nd excited
modes in $A_\mu^{1,2}$ and are referred to as the $W'$ bosons. A
slightly heavier mode comes from the 2nd excited mode in $A_\mu^3$; we
call it the $Z'$ boson. The numerical analysis shows that the masses
of higher Kaluza-Klein modes including the $W'$ and $Z'$ bosons are
around the compactification scale $M_K$ and hierarchically larger than
the SM gauge boson masses (Fig.~\ref{fig:spectrum}). For example, we
have $m_{W'}^{}\,\simeq\, m_{Z'}^{}\,\simeq\,0.83\,(0.82)\,M_K\,
\simeq\,15\,(22)\,m_Z^{}$ for $z_L^{}=10^5\,(10^6)$.
The overall $U(1)$ gauge field $A^4_\mu$ is irrelevant to the
electroweak gauge symmetry and we will not consider Kaluza-Klein modes
from $A^4_\mu$ in the following discussion. The wavefunctions of heavy
gauge bosons $X_\mu^{\alpha(n)}$ are found to be localized at $z=0$
which indicates, from the gauge/gravity correspondence, that these
fields are interpreted as composites (technimesons) in the technicolor
theory. The couplings of $X^{\alpha(n)}_\mu$ bosons to the SM sector
are generally suppressed since their wavefunctions are localized
around $z=0$ and have small overlap with those of the electroweak
gauge bosons. For example the triple gauge boson coupling between $Z$
and $W'$ is evaluated as
\begin{align}
  g_5c_W^{-1} \int dz\, K(z)^\frac{-1}{3} \psi_{W'}(z)^2\psi_Z(z)
  \;\sim\; 0.34\,g
\end{align}
for $z_L^{}=10^5$ (in fact, somehow independently of $z_{L,R}^{}$).
In this way, the above discussion shows that in the dual description
the dynamical electroweak symmetry breaking through the techniquark
condensation is holographically realized. The observed value of the
Weinberg angle can be obtained by taking 
$z_L^{}/z_R^{}\simeq\tan^6\theta_W$ and the $SU(2)$ weak gauge
coupling by choosing $g_5\simeq\sqrt{3}z_L^{1/6} g$.

\medskip

The decay constant $f_{TC}$, which is an analogue of the pion decay
constant of QCD in the technicolor theory, can be calculated in a
similar way to Ref.~\cite{SS}. The Nambu-Goldstone bosons, which are
eaten by the $W$ and $Z$ bosons, are originated from $A_z^{1,2,3}$ and
have the wavefunction proportional to $\partial_z\psi_{0L}(z)$ 
[$=-\partial_z\psi_{0R}(z)$]. Since this wavefunction is localized at
$z=0$, the value of decay constant does not depend on $z_{L,R}^{}$ and
is same as that in Ref.~\cite{SS}:
\begin{align}
  f_{TC}^{} \,=&\; \frac{2}{\sqrt{\pi}g_5}M_K
  \;=\; \frac{kN_{TC}g_{TC}^{}}{3\sqrt{3}\pi^2}\,M_K ,
  \label{f}
\end{align}
where the last equation is obtained from \eqref{g5}. Using this decay
constant we can express the mass spectrum as
\begin{alignat}{3}
  m_Z^2 \,=&\; m_W^2 c_W^{-2},
  &\hspace{8ex}
  m_W^2 \,=&\; \frac{3\pi}{4}\rho_W^{}g^2f_{TC}^2 ,
  &\hspace{8ex}
  m_{X^{(n)}}^2 \,=&\; \lambda_nM_K^2,
 \label{ff}
\end{alignat}
with $\lambda_n\gtrsim{\cal O}(1)$. The last equation suggests the
dynamical scale of technicolor theory is around $M_K$. On the other
hand, the masses of $W$, $Z$ bosons and composites fields (denoted 
by $X$) are estimated from the technicolor theory that
\begin{gather}
  m_Z^2 \,\sim\, \frac{1}{4}(g^2+g'^2) f_{TC}^2,
  \hspace{6ex}
  m_W^2 \,\sim\, \frac{1}{2}g^2 f_{TC}^2,
  \hspace{6ex}
  m_X^2 \,\sim\, \Lambda_{TC}^2, \hspace{6ex} \\[1mm]
  \qquad f_{TC} \,\sim\, \sqrt{N_{TC}}\,\Lambda_{TC} .
\end{gather}From the consistency of these two expressions of spectrum,
we find that the holographic gravity dual provides a calculable and
compatible framework to technicolor theory.

The decay constant $f_{TC}$ is given in terms of $z_L^{}$ and $M_K$. 
It is noticed that the holographic description is valid when the
't~Hooft coupling is large, which might give a constraint on $f_{TC}$
through eq.~\eqref{f}. For example, a large 't~Hooft coupling
$N_{TC}g_{TC}^2=4\pi$ leads to $f_{TC}\simeq0.07k\sqrt{N_{TC}}M_K$. 
When $N_{TC}=10$ and $k=1$ as an example, one obtains
$m_W^{}\sim0.07M_K$. If another condition 
($N_{TC}g_{TC}^2\ll g_{TC}^{-4}$) is taken into account, we would have
a slightly severe constraint on the decay constant.

\smallskip

Finally, several comments are in order. One may wonder about the
unitarity. The general argument in \cite{higgsless} can be applied to
our model as well and then the unitarity of massive gauge theory is
formally recovered by Kaluza-Klein gauge bosons
$X_\mu^{\alpha(n)}$. To avoid the breakdown of perturbative unitarity,
the compactification scale $M_K$ would be set below a few TeV and
$z_L\lesssim{\cal O}(10^7)$. In addition to technimesons, there are
also technibaryons. Refs.~\cite{baryon} study baryons from holographic
descriptions of QCD and the first reference in Refs.~\cite{baryon}
shows that baryons are heavier than the $\rho$ meson after taking into
account the Chern-Simons term, as expected. If we apply their analyses to
our case, we would find technibaryons are heavier than the $W'$ boson.

\bigskip
\section{The Matter Sector}
\label{sec:fl}
\subsection{D-brane configuration}
\label{subsec:matterD}

To complete the realization of the SM (without the Higgs), we next
consider the introduction of SM matter fields. Let us add some number
of D4-branes into the previous brane configuration for the gauge
sector (see Fig.~\ref{fig:brane}) in the flat space. The added
branes are parallel to but separated from the technicolor D4-branes.
In this work these additional D4-branes are referred to as the flavor
branes. Then at the intersection of a flavor D4-brane with D8 
or $\overline{\rm D8}$, we have a massless chiral or anti-chiral
fermion which transforms as the fundamental representation under
$U(N_f)_L$ or $U(N_f)_R$. With appropriate numbers of flavor branes
being introduced, such chiral fermions are identified with the SM
matter fields. There are also open strings which connect the flavor
and technicolor branes. The fermion mass terms are generated by
massive gauge fields from these open strings, which is a similar
mechanism in the extended technicolor theory.

We introduce one flavor D4-brane at one point (for a lepton) and 
three coincident D4-branes at another place (for a quark). These
D4-branes and the technicolor D4-branes are separated to each other
in the extra dimensions, particularly in the $z$ direction. At the
four intersection points among D4'$_{\rm lepton}$, D4'$_{\rm quark}$,
D8 and $\overline{\rm D8}$, we have four types of chiral fermions,
$\ell_L$, $\ell_R$, $q_L^{}$ and $q_R^{}\,$: 
\begin{align*}
\begin{array}{c|cc|cc}
  & SU(2)_L & U(1)_Y & U(1)_l & U(3)_b \\ \hline
  \ell_L & \Box & & 1 & \\
  \ell_R & & (\frac{1}{2},\frac{-1}{2}) & 1 & \\
  q_L^{} & \Box & & & \Box \\
  q_R^{} & & (\frac{1}{2},\frac{-1}{2}) & & \Box 
\end{array}
\end{align*}
where $U(1)_\ell$ and $U(3)_b$ are the gauge symmetries on the
D4'$_{\rm lepton}$ and D4'$_{\rm quark}$ branes, respectively.
Similar to the technicolor branes, the effective theories on the
flavor D4-branes are pure Yang-Mills theories, since scalar and spinor
fields on the flavor branes become massive due to the anti-periodic
boundary conditions for spinors imposed along the $S^1$ direction
($x_4$). Naively the leptons and quarks do not have the correct
hypercharges, but we can mix $U(1)_Y$ with $U(1)_\ell$ and also with
the overall $U(1)_b$ in $U(3)_b$, which are identified to the lepton
and baryon number gauge symmetries. The quark fields $q_{L,R}^{}$ are
assigned to have the $U(1)_b$ charge $1/3$. The mixing depends on how
the $U(1)$ symmetries are broken down. In this work we assume, for
simplicity, that there are two scalar fields at the intersections of
flavor branes and the $\overline{\rm D8}$-branes, whose quantum
numbers are respectively given by $(\frac{1}{2},Q_\ell)$ 
under $U(1)_Y\times U(1)_\ell$ and $(\frac{1}{2},Q_b)$
under $U(1)_Y\times U(1)_b$, and their vacuum expectation values are
taken infinity. In this case the gauge fields $L_\mu$ of $U(1)_\ell$
and $B_\mu$ of $U(1)_b$ at the intersections become
\begin{align}
  L_\mu \,=\, -\frac{g'}{2Q_\ell }Y_\mu ,
  \hspace{8ex}
  B_\mu \,=\, -\frac{g'}{2Q_b}Y_\mu ,
  \label{mix}
\end{align}
where $Y_\mu$ is the gauge fields of $U(1)_Y$ on 
the $\overline{\rm D8}$ branes, and the normalization of $L_\mu$
and $B_\mu$ are taken such that the
gauge couplings appear in front of the kinetic terms.
We then find that the leptons
and quarks have the correct hypercharges with taking a simple choice
$Q_\ell=-Q_b=1$. For example the right-handed quarks $q_R^{}$ have the
minimal interaction with the following combination of gauge fields
\begin{align}
  \frac{\pm 1}{2}ig' Y_\mu +\frac{1}{3}iB_\mu
  \;=\; ig' Y_\mu\times \Big(\frac{2}{3} \mbox{ or } \frac{-1}{3}\Big).
\end{align}
To summarize, the leptons and quarks have charges under the unbroken
gauge symmetry as
\begin{align*}
\begin{array}{c|cc|c}
  & SU(2)_L & U(1)_Y & SU(3)_b \\ \hline
  \ell_L & \Box & \frac{-1}{2} & \\
  \ell_R & & (0,-1) & \\
  q_L^{} & \Box & \frac{1}{6} & \Box \\
  q_R^{} & & (\frac{2}{3},\frac{-1}{3}) & \Box 
\end{array}
\end{align*}
This is just the SM matter content in one generation. To realize the
complete set of three generations, one may further introduce two more
sets of flavor D4-branes and repeat the same mixing. It is assumed
that $SU(3)$'s are broken down to the diagonal $SU(3)_C$ which is
identified to the color gauge symmetry in the SM\@. The 
original $U(1)_Y$ gauge coupling is shifted by the mixing and is
finally matched to the experimentally observed value.

\bigskip
\subsection{Holographic dual description}

We have explained how the SM matter fields are introduced in the
technicolor theory from a viewpoint of brane configuration. Its
holographic description completes a dual picture of electroweak theory
with symmetry breaking caused by strongly-coupled gauge dynamics.

A flavor D4-brane in the near horizon geometry is located at a
constant distance away from the origin $z=0$ and extends along the
$x_4$ direction as well as the non-compact four-dimensional space. The
$i$-th flavor D4-brane intersects with the probe D8-brane at two
points $(x_4,z)=(0,\pm z_i^{})\,$ ($z_i^{}>0$). A left (right) handed
chiral fermion is located at $z=z_i^{}\,$ ($z=-z_i^{}$). Their quantum
charges have been fixed in Section~\ref{subsec:matterD}.

We have presented the scheme in the technicolor side that $U(1)_Y$ on
the $\overline{\rm D8}$ brane is mixed with $U(1)$'s on flavor branes
to have the right hypercharge assignment. The holographic description
of the mixing \eqref{mix} is simply given by replacing $g'$ and
$Y_\mu$ with $g_5$ and $A_\mu^3(-z_i^{})$, i.e.
\begin{align}
  \qquad 
  A^{(i)}_\mu(x,x_4=0) \;=\; -\frac{g_5}{2Q} A^3_\mu(x,-z_i^{}),
  \hspace{6ex}
  (Q=Q_\ell \mbox{ or } Q_b)
  \label{AiA3}
\end{align}
where $A^{(i)}_\mu$ is the gauge field on flavor D4-branes. We have
taken the normalization for $A_\mu^{(i)}(x,\tau)$ such that the gauge
coupling appears in front of the kinetic term. Since the $x_4$
direction is compactified on $S^1$ and gauge fields have periodic
boundary conditions, $A_\mu^{(i)}(x,x_4)$ have a constant profile
along the flavor D4-branes. The photon and $Z$ boson (and the excited
modes $X^{3(n)}_\mu$) are united in $A_\mu^3$ and thus propagate on
the flavor D4-branes. These corrections have the same implication in
the technicolor side where the gauge coupling of $U(1)_Y$ is
shifted. From \eqref{A3}, we obtain
\begin{align}
  A^{(i)}_\mu(x,x_4=0) \;\simeq\; \frac{-g_5s_W^{}}{2\sqrt{3} Qc_W^{}} 
  z_L^\frac{-1}{6} \Big[c_W^{} Q_\mu(x) 
  -\psi_{0R}(-z_i^{}) s_W^{}Z_\mu(x)\Big] +\cdots,
  \label{yy}
\end{align}
where the ellipses denote the terms with $X^{3(n)}_\mu$ bosons. As
long as $z_i$ is large enough, $\psi_{0R}(-z_i^{})$ is almost equal to
one ($\psi_{0R}(-z_i^{})\simeq1+\frac{1}{\pi z_i^{}}$), which implies
from \eqref{yy} that $A^{(i)}_\mu$ is essentially the $U(1)_Y$ gauge
boson. In this case the flavor D4-brane action induces an additional
kinetic term just for $U(1)_Y$ gauge boson and changes the $U(1)_Y$
gauge coupling in canonically normalizing the gauge field. The
parameter $z_R^{}$ is properly adjusted so that the final $U(1)_Y$
gauge coupling constant matches with the observed 
value [see eq.~\eqref{u1yg}]\footnote{This contribution is not
included in the following analysis. A rough estimation shows that the
order of magnitude of $z_R^{}$ is shifted by about one if the gauge
coupling on flavor D4-branes is of the same order of $g$. The
contribution to oblique correction parameters will be discussed in the
next section.}.

Now we are ready to write down the four-dimensional action for the SM
matter fields in the holographic description. We only discuss the
left-handed leptons which emerge from the intersection between the
connected D8-branes and flavor D4-branes. It is a straightforward
extension to include all the matter fields in a completely parallel
way. The action of left-handed leptons becomes
\begin{align}
  S_L \,=&\, \int d^4x\; \bar\ell_L i\gamma^\mu \Big[\partial_\mu 
  -ig_5\frac{\sigma^a}{2}A_\mu^a(x,z_i^{}) -iA^{(i)}_\mu(x,x_4=0) 
  \Big] \ell_L
  \nonumber \\[1mm]
  \,=&\, \int d^4x\; \bar\ell_L i\gamma^\mu \Big[\partial_\mu
  -i\frac{g}{\sqrt{2}}(W_\mu\sigma^++{\rm h.c.}) 
  -ie\Big(\frac{\sigma^3}{2}-\frac{1}{2}\Big)Q_\mu
  -\frac{ie}{2\bar s_f\bar c_f} Z_\mu (g_V^{}-g_A^{}\gamma^5) 
  \Big] \ell_L 
  \nonumber \\[2mm]
  &\hspace{15ex}
  +\big(\mbox{couplings to } X_\mu^{(n)}\big),
  \label{SL}
\end{align}
where $\sigma^+=(\sigma^1+i\sigma^2)/2$, and $\gamma^5=-1$ for
left-handed fermions. The 
parameter $\theta_f$ ($\bar s_f\equiv\sin\theta_f$ 
and $\bar c_f\equiv\cos\theta_f$) is used for denoting the effective
angle to distinguish it from our parametrization $\theta_W$ introduced
in \eqref{wa}. The definition $e=g_5\psi_Q$ is exactly same 
as \eqref{e-gauge} from the gauge boson self coupling. The other
fermion current couplings are defined from the gauge boson
wavefunctions
\begin{align}
  g \,\equiv&\;\, g_5 \psi_W(z_i), \\[2mm] 
  g_A^{} \equiv&\;\, \frac{\bar s_fg_5 \psi_Z(z_i)}{\bar c_fe}
  \cdot \frac{\sigma^3}{2},
  \\[1mm]
  g_V^{} \equiv&\;\, \frac{\bar s_fg_5 \psi_Z(z_i)}{\bar c_fe} 
  \bigg[\frac{\sigma^3}{2}-2\bar s_f \big(\frac{\sigma^3}{2} 
  +\frac{\bar{c}_{f}^2\psi_Z(-z_i)}{2\bar{s}_{f}^2\psi_Z(z_i)}
 \Big)\bigg]. 
\end{align}
The weak gauge couplings seem to depend on the position of flavor
D4-branes. However the gauge boson wavefunctions are nearly constant
in the large $z$ region and the flavor universality of electroweak
gauge coupling is satisfied with good accuracy unless the flavor
branes reside close to the origin $z=0$. The constant profiles of
gauge boson wavefunctions also imply that the electroweak couplings
are approximately given by
$g\simeq g_{WWZ}^{}$, $\;g_A^{}\simeq\frac{\sigma^3}{2}$, and 
$g_V^{}\simeq\frac{\sigma^3}{2}-2\bar s_f^2 
(\frac{\sigma^3}{2}-\frac{1}{2})$, which is consistent with the SM
expressions. We thus find that all the SM matter fields couple to the
photon, $W$ and $Z$ bosons with the (almost) correct strength. On the
other hand, the couplings to higher Kaluza-Klein gauge 
bosons $X_\mu^{a(n)}$ are suppressed because their wavefunctions are
localized at $z=0$ unlike the SM gauge fields. We will show the
numerical results for these behaviors in Section~\ref{sec:STU}.

\bigskip
\subsection{Fermion masses}

Finally we have a brief comment on a possibility how the masses of
matter fermions are generated in the present model. In the extended
technicolor theory, the massive gauge bosons associated with the
breaking of extended technicolor gauge symmetry mediate the
condensation of techniquarks to the SM fermions, leading to their
masses $m_f\sim g_{ETC}^2\langle \bar{Q}_RQ_L\rangle/m_{ETC}^2$ where
$g_{ETC}^{}$ and $m_{ETC}^{}$ are the gauge coupling constant and the
mass of gauge bosons in the extended technicolor theory.

In our D-brane configuration, we have such massive gauge bosons which
originate from open strings stretching between the technicolor D4 and 
flavor branes. That is seen from the fact that the gauge symmetry
is enhanced when the flavor branes attach with the technicolor
branes. The gauge boson mass $m_{ETC}^{}$, which is given by the
length of an open string, and the induced fermion mass are evaluated as
\begin{gather}
  m_{ETC}^{} \;=\; l_s^{-2} \int_0^{z_i} dz\sqrt{-\det g_{\rm os}}
  \;\sim\; \frac{2}{9}N_{TC}g_{TC}^2z_i^\frac{2}{3}M_K ,
  \\[2mm]
  m_q \;\sim\; \frac{81}{4}\frac{g_{ETC}^2}{N_{TC}g_{TC}^4} 
  z_i^\frac{-4}{3} M_K , 
\end{gather}
where $g_{\rm os}$ is the induced metric on the open string which is
localized at a constant $x_4$. For example, if $M_K\sim$ TeV and
$g_{TC}^{}\simeq g_{ETC}^{}$ are assumed, the flavor branes are
located at $z_i^{}\simeq (10, 10^{2.5} ,10^{4,5})$ for the top, charm,
and up quarks, respectively. The positions of flavor branes are within
the cutoff in the $z$ direction. For a small value of $z_{\rm top}$, 
the flavor gauge boson deviates from $U(1)_Y$ and the oblique
correction parameters may be induced. If one may try to cure this
problem, 
an idea is to realize
that the top flavor D4-brane is
not parallel to and has some angle against the technicolor
D4-brane. In this case, a left-handed fermion can be localized closer
to the technicolor brane compared with a right-handed fermion, and one
may obtain a heavy fermion mass without leading to large oblique
correction parameters. However too close to the origin $z\simeq0$, the
wavefunctions for $W$ and $Z$ bosons are deviated from the constant
profiles, and a closer top brane implies that the model would receive
a constraint from the measurement of $Zb_L\bar b_L$
coupling~\cite{PDG}. In addition if one considers the generation
mixing, the rare observation of flavor-changing neutral current would
provide severer constraints.

\bigskip
\section{Electroweak Precision Tests}
\label{sec:STU}

We have constructed a model of electroweak symmetry breaking,
holographically dual to a technicolor theory. It is well known that a
technicolor theory usually suffers from the difficulty of passing the
electroweak precision tests. Any departure from the SM predictions is
severely constrained from the existing experimental data. In
particular, for the electroweak gauge symmetry breaking, that is known
to be summarized as the oblique correction 
parameters; $S$, $T$, $U$~\cite{oblique} which are defined by the
two-point correlation functions of electroweak gauge bosons, and the
vertex correction parameters~\cite{oblique_general}. In this section,
we discuss the tree-level (pseudo) oblique corrections in our
holographic technicolor model.

There are four fundamental parameters in the technicolor theory; the
electroweak gauge couplings $g$ and $g'$, the technicolor scale
$\Lambda$, and the decay constant $f_{TC}$, which correspond to
$z_L^{}$, $z_R^{}$, $M_K$, $g_5$ in the holographic dual
description. The fine structure constant $\alpha$ and the $Z$ boson
mass $m_Z^{}$ are well measured quantities and 
their values at $Z$ pole 
are used to fix $M_K$ and $g_5$.
Therefore we compare our action with the SM action (minus Higgs
fields) at $Z$ pole and parametrize the deviations in the couplings as
oblique parameters. We
obtain four predictions $g$, $g_V^{}$, $g_A^{}$,
and $m_W^{}$ as the functions of $z_L^{}$, $z_R^{}$ and $z_i^{}$.
In the technicolor theory, the predictions are the functions of
$g_{TC}N_{TC}$ and one combination of gauge couplings
[see eqs. \eqref{f} and \eqref{ff}]. 
 The
charged and neutral current interactions and the $W$ boson mass
contain four oblique correction parameters $S$, $T$, $U$, and
$\Delta=\Delta_e+\Delta_\mu$ defined in~\cite{oblique_general}. Since
in the holographic description the gauge fields have been set to
orthonormal, the oblique parameters are expressed in terms of gauge
vertices of SM fermions. From the matter action \eqref{SL}, we find
the following forms of oblique corrections:
\begin{eqnarray}
  \alpha S &=& 4s_{M_Z}^2c_{M_Z}^2\delta_Z
  +4s_{M_Z}^2c_{M_Z}^2\delta_\gamma,
  \\[1mm]
  \alpha T &=& \delta_\rho-2\delta_W+2c_{M_Z}^2\delta_Z
  +2s_{M_Z}^2\delta_\gamma, 
  \\[1mm]
  \alpha U &=& 8s_{M_Z}^2\delta_W -8s_{M_Z}^2c_{M_Z}^2\delta_Z, 
  \\[1mm]
  \Delta &=& \delta_\rho-2\delta_W,
\end{eqnarray}
with the deviations $\delta_{W,Z,\gamma,\rho}$ from the SM form
\begin{gather}
  \delta_W \;\equiv\; 
  \frac{s_{M_Z}^{}\psi_W(z_i^{})}{\psi_Q} -1, \qquad
  \delta_Z \;\equiv\;
  \frac{s_{M_Z}^{}\psi_Z(z_i^{})}{c_{M_Z}^{}\psi_Q}-1, \qquad
  \delta_\gamma \;\equiv\; 
  \frac{-c_{M_Z}^{}\psi_Z(-z_i^{})}{s_{M_Z}^{}\psi_Q}-1, 
  \nonumber \\
  \delta_\rho \;\equiv\; \frac{m_W^2}{m_Z^2c_{M_Z}^2}-1, \qquad
\end{gather}
and $\alpha=1/128.91$ and $s_{M_Z}^2=0.23108$ at the $Z$ pole~\cite{PDG}.
The effective angle $\theta_f$ has been replaced with $\theta_{M_Z}$
which is defined from the Fermi constant. The observed values of these
two angles are almost equal and the difference does not affect the
following analysis. Substituting the approximate solutions obtained in
Section~\ref{sec:DSB}, one has $S=T=U=\Delta=0\,$ if 
$s_{M_Z}^{}=s_W^{}$ is satisfied.

\begin{figure}[t]
\begin{center}
\begin{minipage}{7.5cm}
\begin{center}
\includegraphics[width=8cm]{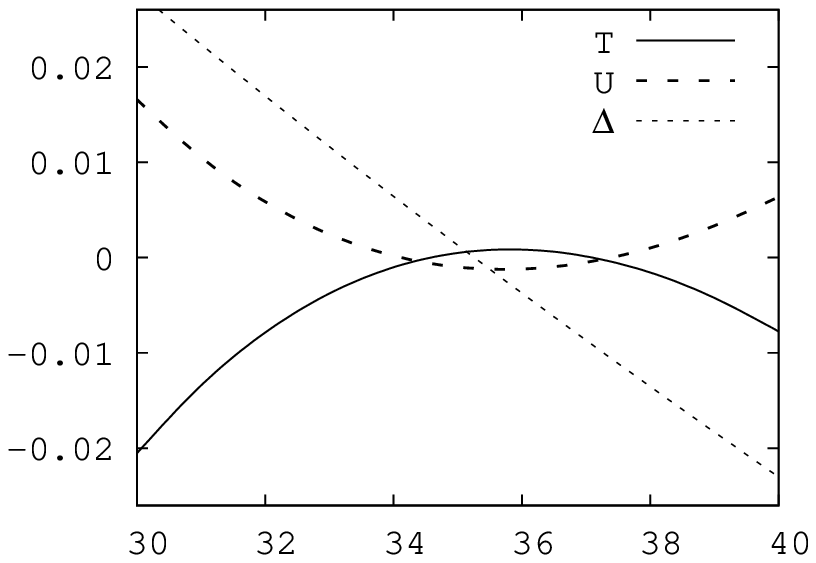}
\put(-108,-6){$z_R^{}/z_L^{}$}
\end{center} 
\end{minipage}
\hspace{7ex}
\begin{minipage}{7.5cm}
\begin{center}\hspace*{-5ex}
\includegraphics[width=8cm]{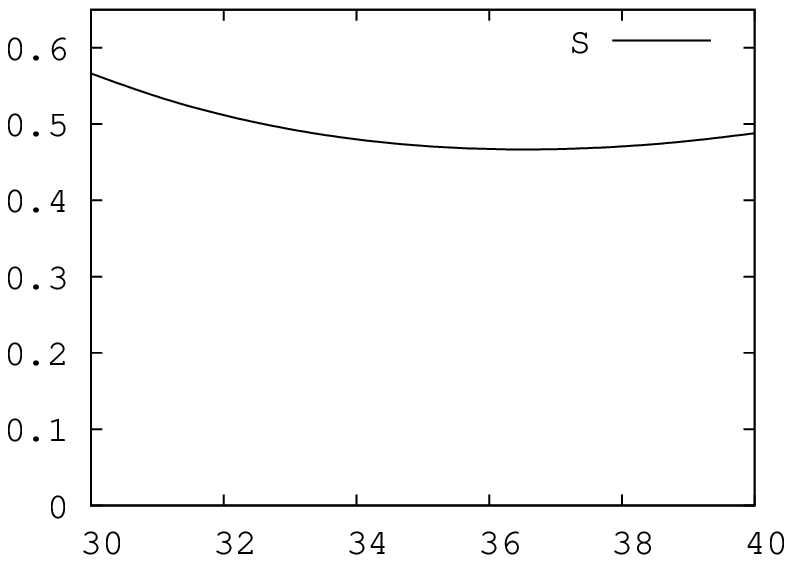}
\put(-111,-6){$z_R^{}/z_L^{}$}
\end{center}
\end{minipage}\smallskip
\caption{The numerical result of oblique correction parameters as the
functions of $z_R^{}$ with $z_L^{}=z_i^{}=10^6$. The fine structure
constant and the electroweak gauge boson mass are fitted to the
experimental data.\bigskip}
\label{fig:stud}
\end{center}
\end{figure}

Let us first study the case that all the SM matter fields are
localized at the same point $z_i^{}=z_L^{}$ (and can be separated
along $S^4$ direction) and discuss the effects of changing the
position of flavor branes later. The numerical results are summarized
in Table~\ref{table:stu} and Fig.~\ref{fig:stud}.
\begin{table}[t]
\begin{center}
$\begin{array}{c|c|ccc|ccc|ccc}
  z_L^{} & z_R^{}/z_L^{} & g & g_V^\ell & g_A^\ell
  & g_{WWZ}^{} & g_{WWZZ}^{} & g_{WWWW}^{} & S & T & U \\ \hline\hline
  10^4 & 29.81 & 0.644 & -0.0263 & 0.494
  & 0.977 & 0.958 & 0.967 & \,2.26\, & 0.011 & -0.025 \\ \hline
  10^5 & 33.50 & 0.647 & -0.0329 & 0.498
  & 0.990 & 0.981 & 0.985 & 1.02 & 0.003 & -0.005 \\ \hline
  10^6 & 35.26 & 0.648 & -0.0356 & 0.499
  & 0.995 & 0.991 & 0.993 & 0.47 & \sim0 & -0.001 \\ \hline
  10^7 & 36.10 & 0.649 & -0.0368 & 0.499
  & 0.998 & 0.996 & 0.997 & 0.22 & \sim0 & \sim0 
\end{array}$\medskip
\caption{The numerical result of the tree-level oblique correction
parameters. The coupling constants of triple and quartic gauge bosons
are shown as the fractions to the SM expressions. The $\Delta$
parameter is chosen to be zero in the table. The fine structure
constant and the $Z$ boson mass are fitted to the experimental
data.\bigskip}
\label{table:stu}
\end{center}
\end{table}
Table~\ref{table:stu} shows the $S$, $T$, $U$ parameters with
$\Delta=0$ which is fixed by choosing an appropriate value of
$z_R^{}/z_L^{}$.
In Fig.~\ref{fig:stud} the oblique parameters are displayed as the
functions of $z_R^{}/z_L^{}$. We have set $z_L^{}=10^6$ from the
hierarchy between the electroweak and Kaluza-Klein excited gauge
bosons. From these results, we find that the oblique parameters except
for $S$ are roughly constant and small compared with the SM fit: 
$S=-0.13\pm 0.10$, $T=-0.13\pm 0.11$ and 
$U= 0.20\pm0.12$~\cite{PDG} (while one could subtract the contribution
of Higgs fields). The smallness of $T$ parameter is ensured because of
the custodial symmetry. The $S$ parameter is generally large and
positive, but decreases as $z_L^{}$ to the experimentally allowed
region for $z_L^{}\gtrsim{\cal O}(10^7)$.
For $z_L\gtrsim 10^7$, $N_{TC}g_{TC}\lesssim 4$ with assuming 
$k\simeq 1$, and so the validity of holography may not be clear.
Table~\ref{table:stu} also
shows that the self couplings of gauge bosons are consistent with the
observed data. Such a result has been mentioned in the previous
section with the approximate solutions. The holographic dual
description thus recovers the qualitative behavior of technicolor
theory against the electroweak precision test. However the holographic
theory has some advantages that the oblique correction parameters are
easily handled by deforming the model and/or taking other sources to
the corrections into account.

In the evaluation of oblique correction parameters, there are other
sources, than $z_R^{}/z_L^{}$ shown above, which lead to the
modification of oblique parameters, in particular, the reduction of
$S$ parameter. The first is the position of SM fermions in the extra
dimensions, i.e.\ the intersecting point of D8 and flavor D4 branes. 
The position of flavor branes little affects the tree-level oblique
parameters in a large $z$ region because the wavefunctions of
electroweak massive gauge bosons have almost flat profiles along the
extra dimension. If one places the flavor branes at some point closer
to the technicolor branes, the $S$ parameter is reduced and can be
negative, since the fermion couplings to the $Z$ boson is a bit
suppressed. For example, $S=-0.056$, $T=-0.267$, $U\sim0$ and
$\Delta=0.002\,$ for $z_i^{}=300$ and $z_L^{}=10^6$. The second
possible source is the contribution from the flavor branes. If the
gauge field on the flavor D4-brane is just proportional to $U(1)_Y$,
an additional kinetic term is absorbed by changing the $U(1)_Y$ gauge
coupling $g'$, i.e.\ by adjusting $z_R^{}$. The non-vanishing oblique
corrections are induced when the gauge group on the flavor branes
differs from $U(1)_Y$, that is, 
if $\Gamma\equiv c_W^{}\psi_Z(-z_i)^{}/(s_W^{}\psi_Q)$ is different
from $-1$. From the numerical analysis, we find 
$\Gamma=-1.007$ ($-1.003$) for $z_i^{}=10^6$ ($300$) 
and $z_L^{}=10^6$. That induces an extra kinetic term for the $Z$
boson and the $S$ parameter is pushed toward negative with an amount
of $\propto g^2(\Gamma+1)/g_i^2$ whose size depends on the gauge
coupling constant $g_i$ on the flavor D4-branes.

We have also not included the corrections from
technimesons. Table~\ref{table:KK} shows the numerical evaluation for
the masses and coupling constants of Kaluza-Klein excited gauge bosons.
\begin{table}[t]
\begin{center}
$\begin{array}{c|c|ccc|cccc}
  z_L^{} & z_R^{}/z_L^{} & M_K & M_{W'} & M_{Z'}
  & g_{W'W'Z}^{} & g_{WWZ'}^{} & g_{WWZ''} 
  & g_{ffW'}^{} \\ \hline\hline
  10^4 & 29.81 & 1100 & 917  & 923  & 0.323 & 0.0549 & 
  0.000440 & 0.192 \\ \hline
  10^5 & 33.50 & 1646 & 1359 & 1362 & 0.338 & 0.0378 & 
  0.000147 & 0.132 \\ \hline
  10^6 & 35.26 & 2437 & 2002 & 2004 & 0.344 & 0.0259 & 
  0.000048 & 0.090 \\ \hline
  10^7 & 36.10 & 3591 & 2943 & 2945 & 0.347 & 0.0177 &
  0.000015 & 0.062
\end{array}$
\end{center}
\caption{The masses and coupling constants of Kaluza-Klein excited
gauge bosons. The gauge bosons $W'$, $Z'$ and $Z''$ are the 2nd and
3rd excited Kaluza-Klein modes in $A_\mu^{1,2,3}$. The mass parameters
are denoted in GeV unit and the higher-mode couplings are given by the
ratio to the corresponding SM couplings.\bigskip}
\label{table:KK}
\end{table}
The higher-mode gauge couplings are expressed by the ratios to the
corresponding SM couplings. It is found from the table that the
tree-level correction to the Fermi constant is 
roughly ${\cal O}(10^{-4})$ and the $T$ parameter is shifted toward
negative with amount of ${\cal O}(10^{-1})$. Moreover the higher
Kaluza-Klein mode couplings to fermions are more suppressed. In a
recent paper~\cite{CES}, a possibility is pointed out that $S$ may be
modified depending on the distance between D8 
and $\overline{\rm D8}$-branes. Among various contributions, which
effect is dominant depends on the model parameters and the details of
calculation is left for future study. A small (and
even negative) $S$ parameter is expected to be viable if taken into
account these modifications of the model.

\bigskip
\section{Comparison to Higgsless Models}
\label{sec:comp}

In this section, we comment on some connections to the so-called
higgsless models which are defined in five
dimensions~\cite{higgsless}. The viable types of higgsless models
utilize the AdS$_5$ warped geometry and the electroweak symmetry
breaking is caused by appropriate boundary conditions at the infrared
(IR) brane. 

\begin{figure}[t]
\begin{center}
\includegraphics[width=7.2cm]{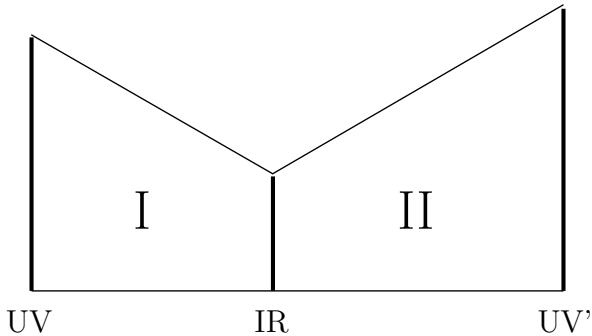}
\put(-165,25){\LARGE I}
\put(-65,25){\LARGE II}
\put(-12,-14){UV'}
\put(-119,-14){IR}
\put(-213,-14){UV}\medskip
\caption{The two throats meet at one IR brane.\bigskip}
\label{fig:twoUV}
\end{center}
\end{figure}

An analogy to higgsless models in the warped geometry becomes that we
have two throats which merge at the IR brane. The two boundaries in
our model correspond to two different ultraviolet (UV) branes (the UV
and UV' branes in Fig.~\ref{fig:twoUV}).
In fact the boundary conditions in our model determine the gauge
symmetry above the technicolor scale. In the language of higgsless
models, this corresponds to that there exist two different gauge
symmetries in the two throat regions, and the boundary conditions at
the two UV branes determine what of symmetries are gauged at UV
scales.

In the higgsless models, the electroweak gauge symmetry is broken
by the boundary conditions imposed at the IR brane. It may be
suggested from the gauge/gravity correspondence in string theory that
a higgsless model in the AdS$_5$ geometry with boundaries has a dual
description in terms of strongly coupled gauge theory. However it is
generally difficult to determine the dynamics of technicolor theory.
On the contrary to that, in our model, we can identify the technicolor
theory and the condensation of techniquarks. That implies in the
higgsless models that a specific boundary condition at the IR brane is
chosen to connect the gauge fields in the bulks I and II.

The limit of taking $z_R^{}\to0$, i.e.\ taking the UV' brane in
Fig.~\ref{fig:twoUV} close to the IR brane, might be thought as the
reduction to a higgsless model. This is however unlikely since the
limit corresponds to a strong coupling limit of the gauge symmetry 
on the $\overline{\rm D8}$ branes.

In the higgsless scenarios, the oblique correction parameters have
been explored in the literature. In particular it was pointed out that 
the $S$ parameter is made small if bulk SM matter fields are
introduced in a specific way~\cite{higgsless2}. The situation is
similar to our model in which the $S$ parameter becomes smaller if we
place the flavor D4-branes closer to the technicolor D4-branes. There
have been various proposals in the higgsless models to reduce oblique
corrections and to avoid a large deviation in the $Zb\bar b$ coupling
while realizing the heavy top quark. These proposals may offer the
suggestions for modifying our model.

\bigskip
\section{Conclusions and Discussions}

In this paper we have explored a holographic dual description of
technicolor theory from the D-brane configuration. The electroweak
gauge symmetry is dynamically broken in the D4 background geometry.
The holographic description makes it possible to analyze the
non-perturbative dynamics of technicolor theory in a perturbative and
quantitative way. We have calculated the spectrum of SM gauge bosons
and technimesons which are expressed by the technicolor scale and the
decay constant. The heavier mode gauge bosons obtain hierarchically
larger masses than the SM ones and have suppressed couplings to the SM
matter fields. The quarks and leptons have been introduced with the
correct hypercharges from the flavor D4-branes and their masses are
generated by massive gauge bosons in a similar way to the extended
technicolor theory.

The oblique parameters have been numerically computed and found to be
small, except for the $S$ parameter which significantly deviates from
zero and takes a positive value. We have discussed several sources to
reduce the $S$ parameter (even toward a negative value): the positions
of flavor branes, and the contribution to hypercharge kinetic terms
from the flavor branes. Another interesting possibility would be to
realize bulk SM fermions. For example, with an additional D8-brane
introduced, an open string between this new D8 and the electroweak D8
branes induces a pair of vector-like quarks. The bulk fermion mass
parameters are tuned by the distance between two D8-branes. The
introduction of bulk fermions would also be useful for reducing
oblique corrections, realizing the correct couplings of the third
generation fermions, suppressing flavor-changing rare processes, and
so on.

It is an independent question if a phenomenologically viable model can
be constructed by D-brane configurations. The fluxes which stabilize
moduli often generate throats and a more realistic model may appear in
other throat geometries. As well as other ways of embedding the
electroweak symmetry and introducing the SM matter, the applications
to higher-scale theory such as grand unified theory would be
worthwhile.

\bigskip\bigskip
\subsection*{Acknowledgements}

The authors would like to thank Kang-Sin~Choi for collaboration during
the early stages of this work. T.H.\ would like to thank 
Koji~Hashimoto and Akitsugu~Miwa for useful discussions. This work is
supported by the European Union 6th framework program
MRTN-CT-2004-503069 ``Quest for unification'',
MRTN-CT-2004-005104 ``ForcesUniverse'',
MRTN-CT-2006-035863 ``UniverseNet'',
SFB-Transregio 33 ``The Dark Universe'' by Deutsche
Forschungsgemeinschaft (DFG), the grant-in-aid for scientific 
research on the priority area (\#441) ``Progress in elementary
particle physics of the 21st century through discoveries of Higgs
boson and supersymmetry'' (No.~16081209) and by scientific grant
from Monbusho (No.~17740150).

\bigskip\bigskip


\begin{thebibliography}{99}

\bibitem{TC}
S.~Weinberg,
Phys.~Rev. {\bf D13} (1976) 974;
Phys.~Rev. {\bf D19} (1979) 1277.

L.~Susskind,
Phys.~Rev. {\bf D20} (1979) 2619.

C.T.~Hill and E.H.~Simmons,
Phys.~Rept. {\bf 381} (2003) 235
[Erratum-ibid. {\bf 390} (2004) 553].


\bibitem{oblique}
B.~Holdom and J.~Terning,
Phys.~Lett. {\bf B247} (1990) 88.

M.E.~Peskin and T.~Takeuchi,
Phys.~Rev.~Lett. {\bf 65} (1990) 964;
Phys.~Rev. {\bf D46} (1992) 381.

M.~Golden and L.~Randall,
Nucl.~Phys. {\bf B361} (1991) 3.

G.~Altarelli and R.~Barbieri,
Phys.~Lett. {\bf B253} (1991) 161.


\bibitem{AdSCFT}
J.M.~Maldacena,
Adv.~Theor.~Math.~Phys. {\bf 2} (1998) 231
[Int.~J.~Theor.~Phys. {\bf 38} (1999) 1113]
[arXiv:hep-th/9711200].


\bibitem{AdSCFT2}
S.S.~Gubser, I.R.~Klebanov and A.M.~Polyakov,
Phys.~Lett. {\bf B428} (1998) 105
[arXiv:hep-th/9802109].

E.~Witten,
Adv.~Theor.~Math.~Phys. {\bf 2} (1998) 253
[arXiv:hep-th/9802150].

O.~Aharony, S.S.~Gubser, J.M.~Maldacena, H.~Ooguri and Y.~Oz,
Phys.~Rept. {\bf 323} (2000) 183
[arXiv:hep-th/9905111].


\bibitem{SS}
T.~Sakai and S.~Sugimoto,
Prog.~Theor.~Phys. {\bf 113} (2005) 843
[arXiv:hep-th/0412141];
{\it ibid.} {\bf 114} (2006) 1083
[arXiv:hep-th/0507073].


\bibitem{ETC}
S.~Dimopoulos and L.~Susskind,
Nucl.~Phys. {\bf B155} (1979) 237.

E.~Eichten and K.D.~Lane,
Phys.~Lett. {\bf B90} (1980) 125.


\bibitem{higgsless}
C.~Csaki, C.~Grojean, H.~Murayama, L.~Pilo and J.~Terning,
Phys.~Rev. {\bf D69} (2004) 055006
[arXiv:hep-ph/0305237].

C.~Csaki, C.~Grojean, L.~Pilo and J.~Terning,
Phys.~Rev.~Lett. {\bf 92} (2004) 101802
[arXiv:hep-ph/0308038].


\bibitem{D4metric}
E.~Witten,
Adv.~Theor.~Math.~Phys. {\bf 2} (1998) 505
[arXiv:hep-th/9803131].


\bibitem{stringQCD}
M.~Kruczenski, D.~Mateos, R.C.~Myers and D.J.~Winters,
{\it ibid.} {\bf 0405} (2004) 041
[arXiv:hep-th/0311270].


\bibitem{baryon}
H.~Hata, T.~Sakai, S.~Sugimoto and S.~Yamato,
arXiv:hep-th/0701280.

D.K.~Hong, M.~Rho, H.U.~Yee and P.~Yi,
arXiv:hep-th/0701276.

K.~Nawa, H.~Suganuma and T.~Kojo,
Phys.\ Rev.\  D {\bf 75} (2007) 086003
[arXiv:hep-th/0612187].

D.~K.~Hong, M.~Rho, H.~U.~Yee and P.~Yi,
arXiv:0705.2632 [hep-th].




\bibitem{oblique_general}
C.P.~Burgess, S.~Godfrey, H.~Konig, D.~London and I.~Maksymyk,
Phys.~Rev. {\bf D49} (1994) 6115
[arXiv:hep-ph/9312291].


\bibitem{PDG}
W.M.~Yao {\it et al.} [Particle Data Group],
J.~Phys. {\bf G33} (2006) 1.


\bibitem{CES}
C.D.~Carone, J.~Erlich and M.~Sher,
arXiv:0704.3084 [hep-th].


\bibitem{higgsless2}
G.~Cacciapaglia, C.~Csaki, C.~Grojean and J.~Terning,
Phys.~Rev. {\bf D71} (2005) 035015
[arXiv:hep-ph/0409126].

R.~Foadi, S.~Gopalakrishna and C.~Schmidt,
Phys.~Lett. {\bf B606} (2005) 157
[arXiv:hep-ph/0409266].


\end{thebibliography}
\end{document}